\documentclass[journal]{IEEEtran}

\usepackage{amsfonts,amsmath,amsthm}
\usepackage{subfigure}
\usepackage{multirow}
\usepackage{psfrag}
\usepackage{pifont}
\usepackage{rotating}
\usepackage{threeparttable}
\usepackage{graphicx}

\newtheorem{thm}{Theorem}

\newtheorem{rem}{Remark}

\newcommand{\RR}{{\mathbb R}}

\newcommand{\CC}{{\mathbb C}}

\newcommand{\dotex}{{\frac{d}{dt}}}

\newcommand{\ket}[1]{\left|#1\right>}
\newcommand{\bra}[1]{\left<#1\right|}
\newcommand{\bket}[1]{\left<#1\right>}
\newcommand{\tr}[1]{\text{Tr}\left(#1\right)}

\begin{document}

\bibliographystyle{plain}

\title{Singular perturbations and Lindblad-Kossakowski differential equations}
%
%
%

\author{Mazyar~Mirrahimi and
        Pierre~Rouchon
\thanks{M. Mirrahimi is with the SISYPHE team, INRIA Rocquencourt, Domaine de Voluceau,
B.P. 105, 78153 Le Chesnay Cedex, France. e-mail: mazyar.mirrahimi@inria.fr}
\thanks{P. Rouchon is with the Centre Automatique et Syst\`emes, Ecole des Mines de Paris, 60 Bd Saint-Michel, 75272 Paris cedex 06, France,
e-mail: pierre.rouchon@ensmp.fr} }

\maketitle

\begin{abstract}
We consider an ensemble of quantum systems whose average evolution
is described by a density matrix,  solution of a
Lindblad-Kossakowski differential equation. We  focus on the special
case where  the decoherence is only due to a highly unstable excited
state and where the spontaneously emitted photons are measured by a
photo-detector. We propose  a systematic method to eliminate  the
fast and asymptotically stable  dynamics associated to the excited
state in order to obtain another differential equation for the slow
part. We show that this slow differential equation is still of
Lindblad-Kossakowski type, that  the decoherence terms and the
measured output  depend  explicitly on the amplitudes of
quasi-resonant applied field, i.e., the control.  Beside  a rigorous
proof of the slow/fast (adiabatic) reduction based on  singular
perturbation theory,  we also provide   a physical  interpretation
of the result in the context of coherence population trapping via
dark states and decoherence-free subspaces. Numerical simulations
illustrate the accuracy of the proposed approximation for a 5-level
systems.
\end{abstract}


\IEEEpeerreviewmaketitle

\section{Introduction}\label{sec:intro}

Under the usual assumptions of  optical pumping  and/or coherent
population trapping,  the  Lindblach-Kossakowski  master equation
describing  the dynamics of the density operator admits  multiple
time-scales. In this paper, we are studying the fast/slow structure
resulting from a separation between
\begin{itemize}
 \item the life-times of the excited and unstable states assumed to be short.
 \item the oscillation periods, assumed to be long,   associated to the energies of
 the other stable states and to the Rabi pulsation generated by the  control,  coupling
 the unstable and stable states in a quasi-resonant way.
\end{itemize}
Usually, the elimination of the fast dynamics is performed  in terms
of coherence vector gathering in a single column the coefficients of
the density matrix. In this form, the system is not written in a
standard form~\cite[chapter 9]{khalil-book} (also called Tikhonov
normal form)  with a clear splitting of the coherence vector into
two sub-vectors: a fast sub-vector  and a slow one. Some tedious
linear algebra and changes of variables  are then  needed to put the
system into the  standard  form in order to perform the adiabatic
(quasi-static approximation) elimination of the fast dynamics.
Moreover  with such coherence vector we loose the physical
interpretation of the master equation in terms of Hamiltonian and
jump operators, explained in~\cite[chapter
4]{haroche-raimond:book06}.

The main contribution of this note is to propose a more intrinsic
elimination of the fast part of the dynamics by using only matrix
manipulations for systems  with a structure sketched on
figure~\ref{fig:model}. The main theoretical guide  is the geometric
theory of singularly perturbed differential systems initiated
in~\cite{fenichel-79} and center manifold techniques to approximate
the invariant slow manifold~\cite{khalil-book,carr-book}. Such
theoretical guides have been already used
in~\cite{duchene-rouchon-ces96} in the context of reduction of
kinetics combustion schemes.  These guides  avoid here  the use of
the coherence vector and provide a slow dynamics  that is also a
Markovian master equation of Lindbald-Kossakowski type with a slow
Hamiltonian and slow  jump operators. This slow master equation
describes  the dynamics of the   density matrix  of the  open
quantum system that lives in the  Hilbert space spanned by the
stable states.  As far as we know, such formulation of the slow
dynamics is new, even in the physicist community, and could be of
some interest for the control. In particular, the controls appear
explicitly  in the decoherence terms and the output map.

The note is organized as follows. Section~\ref{sec:3scales} is
devoted  to the modeling of systems depicted on
figure~\ref{fig:model}, to the three time-scales  structure and  to
the elimination (by averaging, i.e., by the rotating wave
approximation  usually used by physicists) of the fastest time-scale
attached to the transition frequencies between the stable and
unstable states. The resulting model, equation~\eqref{eq:twoscale}
with complex value controls $\Omega_k$ and measured output  $y$,
still admits  two time-scales, an asymptotically stable fast part
and a slow part. Extraction of the slow part  is the object of
section~\ref{sec:adi} where the approximation Theorem~\ref{thm:main}
is  proved. For readers not interested  by these technical
developments, we  have summarized at the end of this
section~\ref{sec:adi}  the main formula for deriving the slow master
equation~\eqref{eq:slowbis} from the  original slow/fast one. In
section~\ref{sec:phys}, we compute the slow approximation  when the
Hamiltonian $H$ corresponds to~\eqref{eq:twoscale} and provides
physical interpretations  in terms of slow Hamiltonian and slow jump
operator depending directly on the control amplitudes $\Omega_k$. In
section~\ref{sec:num}, we compare,  numerically on a five-level
system, the slow/fast master equation  with the slow one.

A preliminary version of these results can be found in~\cite{mirrahimi-rouchon-cdc06}.
The authors thank Guilhem Dubois from LKB  for several discussions on  the physics underlying coherence population trapping.

\section{The three time-scale master equation} \label{sec:3scales}

\begin{figure}[htp]
\begin{center}
\includegraphics[width=.48\textwidth]{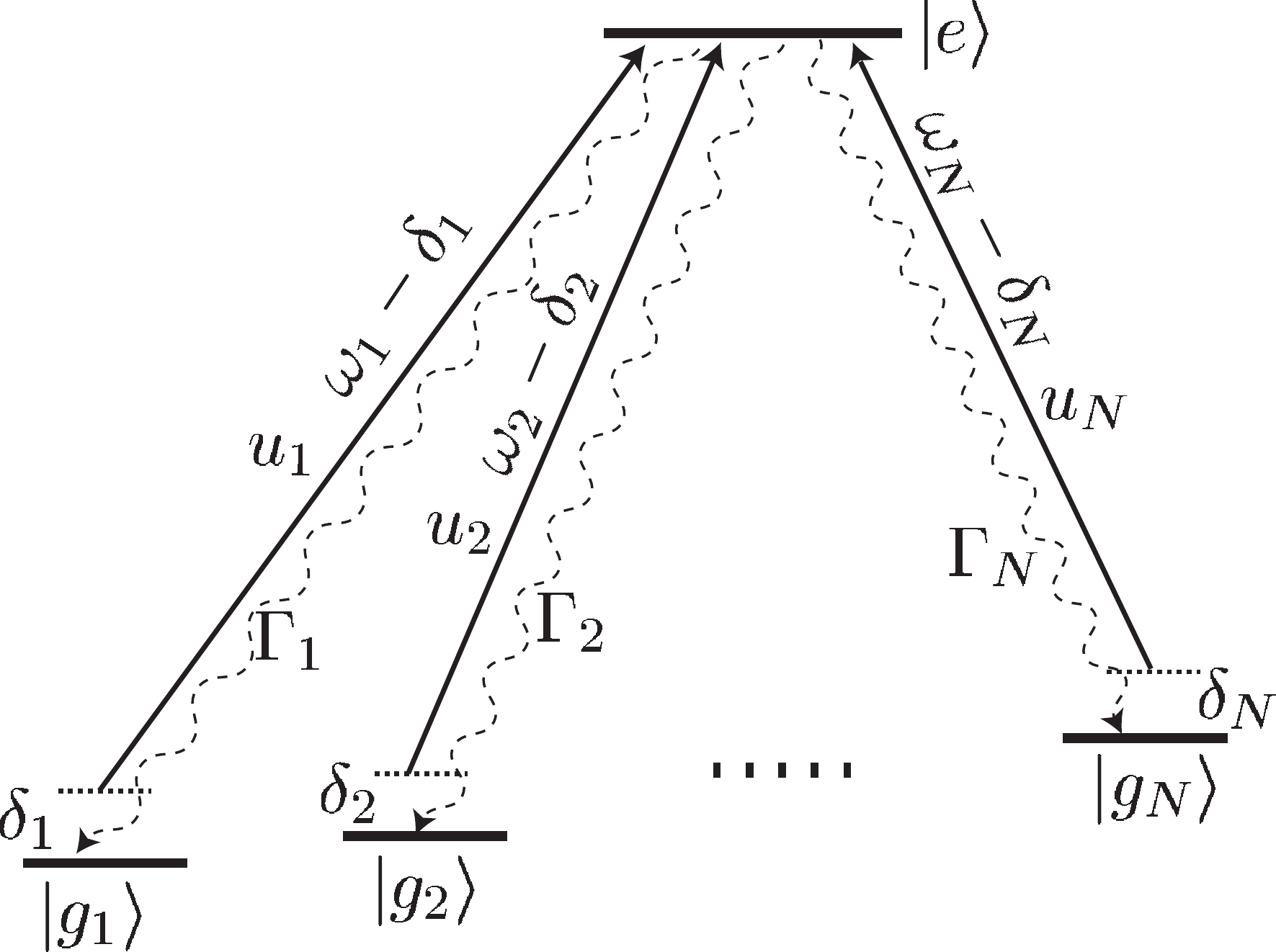}
\caption{relevant energy levels, transitions and decoherence rates
for the considered model~\eqref{eq:model}}\label{fig:model}
\end{center}
\end{figure}

Such master equations  typically describe coherent population
trapping  when a  laser irradiates  an $(N+1)$-level
system~\cite{arimondo-76,arimondo-96}. The system is composed of $N$
(fine or hyperfine) ground states $\left(\ket{g_k}\right)_{k=1}^N$
having energy separations in the radio-frequency or microwave
region, and an excited state $\ket{e}$ coupled to the lower ones  by
optical transitions at frequencies $\left(\omega_k\right)_{k=1}^N$
(see Figure~\ref{fig:model}). Naturally, the decay times for the
optical coherences are assumed to be much shorter than those
corresponding to the ground state transitions (here assumed to be
metastable).

The quantum Markovian master equation of Lindblad-Kossakowski type,
modeling the evolution of a statistical ensemble of identical
systems given by Figure~\ref{fig:model}, reads (see~\cite[chapter
4]{haroche-raimond:book06} for a tutorial and up-to-date  exposure
on such master equations):
\begin{align}
\dotex \rho
  &=-\frac{\imath}{\hbar}[H_0+u H_1,\rho] \notag
  \\&\quad
  +\frac{1}{2}\sum_{k=1}^N  \Gamma_k (2Q_k\rho Q_k^\dag-Q_k^\dag Q_k\rho-\rho Q_k^\dag Q_k) \label{eq:model}
\\
y &= \sum_{k=1}^{N} \Gamma_k \tr{Q_k^\dag Q_k \rho} \label{eq:y}
\end{align}
where $u\in\RR$ is the controlled input (laser field) and $y\geq 0$
is the measured output (number  of photons per time unit
spontaneously emitted from the excited state $\ket e$). Here the
Hermitian operators $H_0$ and $H_1$ are, respectively, the free
Hamiltonian and the interaction Hamiltonian with a coherent source
of photons $u(t)\in \RR$:
\begin{align*}
    \frac{H_0}{\hbar} &=
      \lambda \ket{e}\bra{e} + \sum_{k=1}^{N} \lambda_k \ket{g_k}\bra{g_k}
    \\
    \frac{H_1}{\hbar} &=    \sum_{k=1}^{N} \mu_k (\ket{g_k}\bra{e}+\ket{e}\bra{g_k})
\end{align*}
the $\mu_k$ being coupling  and constant parameters.
Moreover, the quantum jump
operators $Q_k$ corresponding to the spontaneous emission from the
state $\ket{e}$ towards $\ket{g_k}$ are given as follows
$$
Q_k=\ket{g_k}\bra{e}.
$$
One easily has the following relations
\begin{equation}\label{eq:relation}
Q_k Q_l=0,\quad Q_k^\dag Q_k= P=\ket{e}\bra{e} \quad \forall k\neq
l\in\{1,\cdots,N\}.
\end{equation}
The transition frequencies, $\omega_k = \lambda-\lambda_k$ (where
$\lambda$ and $\lambda_k$ are the eigenvalues of $\frac{H_0}{\hbar}$ corresponding
to the energy levels $\ket{e}$ and $\ket{g_k}$, respectively), are
supposed to be much larger than the decoherence rates $\Gamma_k$.
 The  control field $u(t)$ is assumed to be  in the
quasi-resonant regime with respect to the natural frequencies of the
system:
\begin{equation}\label{laser:eq}
u(t)=\sum_{k=1}^N u_k e^{\imath(\omega_k-\delta_k)t} + u_k^\ast e^{-\imath(\omega_k-\delta_k)t}
\end{equation}
where the complex amplitudes $u_k\in\CC $ are varying slowly  and  where $\delta_k$ are the  small de-tuning  frequencies. We  have thus  three different time scales:
\begin{enumerate}
  \item The very fast time-scale associated to the optical frequencies $\omega_k$.
  \item The fast time-scale associated to the life times of the excited state $\ket e$, $\Gamma_k$.
  \item The slow time-scale associated to the laser amplitude $|\mu_k u_k|$ and to the other atomic  transition frequencies $\omega_{kl}= \omega_k-\omega_{l}$, $k\neq l$.
\end{enumerate}
We are interested here  by the  slow time-scale  of system~\eqref{eq:model} where the control $u(t)$ is given by~\eqref{laser:eq} with the following time-scales separation:
$$
 |\omega_{kl}|, |\mu_k u_k| \ll \Gamma_{k^\prime} \ll \omega_{k^{\prime\prime}}
 \quad\text{and}\quad
 \left| \dotex u_k\right| \ll \Gamma_{k^\prime} |u_k|
$$
with $k,l,k^\prime,k^{\prime\prime}\in\{1,\ldots,N\}$, $k\neq l$.

Elimination of the fastest time-scales is standard. It corresponds
to the so-called rotating wave approximation and can be justified by
averaging techniques. This is not the object of this note and thus
we just recall here the resulting master equation:
\begin{equation}\label{eq:RWA}
\frac{d}{dt}\rho=-\frac{\imath}{\hbar}[H,\rho]+\frac{1}{2}\sum_{k=1}^N
\Gamma_k (2Q_k\rho Q_k^\dag-Q_k^\dag Q_k\rho-\rho Q_k^\dag Q_k).
\end{equation}
Calculating the secular terms of $u e^{-\imath\frac{H_0}{\hbar} t} H_1 e^{\imath\frac{H_0}{\hbar} t}$, the effective Hamiltonian $H$ is given as follows:
\begin{equation}\label{eq:H-eff}
\frac{H}{\hbar}=\sum_{k=1}^N \delta_k \ket{g_k}\bra{g_k}+\Omega_k
\ket{g_k}\bra{e}+\Omega_k^* \ket{e}\bra{g_k}.
\end{equation}
with $\Omega_k= \mu_k u_k$.
Note that the measured output remains unchanged
\begin{equation}\label{eq:measure}
y=\left(\sum_k \Gamma_k\right)\tr{P\rho}=\left(\sum_k
\Gamma_k\right)\tr{\ket{e}\bra{e}\rho}.
\end{equation}
We are  led   to the following  master equation
\begin{align}
  \dotex \rho &= -\imath\left[
     \sum_{k=1}^N \delta_k \ket{g_k}\bra{g_k}+\Omega_k
\ket{g_k}\bra{e}+\Omega_k^* \ket{e}\bra{g_k}~ , ~ \rho
    \right]
  \notag \\ & \quad
  +
  \sum_{k=1}^N
\frac{\Gamma_k}{2} \text{\Large (}2\bket{e|\rho|e} \ket{g_k}\bra{g_k} -\ket e\bra e\rho-\rho \ket e\bra e
  \text{\Large )}
  \label{eq:twoscale}
  \\
  y &= \left(\sum_{k=1}^{N}
\Gamma_k\right)\tr{\ket{e}\bra{e}\rho} \notag
\end{align}
where  the $\Omega_k$'s are the slowly varying  complex  amplitudes
(controlled inputs), the $\delta_k$'s are the laser  de-tunings
and where  the two time-scales separation  results from:
$$
|\delta_k|,|\Omega_k| \ll \Gamma_{k^\prime} \quad \text{and}\quad
\left| \dotex \Omega_k \right| \ll \Gamma_{k^\prime} |\Omega_k|
$$
for $k,k^\prime\in\{1,\ldots, N\}$.

\section{Slow/fast reduction}\label{sec:adi}
 We can therefore take
$\Gamma_k=\overline\Gamma_k/\epsilon$ where $\epsilon$ is a small
positive parameter. Thus we have a  master
equation  with the following structure:
\begin{equation}\label{eq:fastdecay}
\frac{d}{dt}\rho=-\frac{\imath}{\hbar}[H,\rho]+\sum_{k=1}^N
\frac{\overline\Gamma_k }{2\epsilon}(2Q_k\rho Q_k^\dag-Q_k^\dag
Q_k\rho-\rho Q_k^\dag Q_k),
\end{equation}
where $\overline \Gamma_k$'s and $H$ (given by~\eqref{eq:H-eff} for example)  have the same orders of
magnitude but where $\epsilon >0$ is a small parameter.

Define, with $P=\ket e \bra e$,
\begin{align}\label{eq:slowfast}
\rho_f &= P\rho+\rho P-P \rho P \notag\\
\rho_s &= (1-P)\rho(1-P)+\frac{1}{\Big(\sum_{k=1}^N
\overline\Gamma_k\Big)}~\sum_{k=1}^N \overline\Gamma_k~Q_k \rho
Q_k^\dag.
\end{align}
We have
\begin{equation}\label{eq:inverse}
\rho=\rho_s+\rho_f-\frac{1}{\Big(\sum_{k=1}^N
\overline\Gamma_k\Big)}~\sum_{k=1}^N \overline\Gamma_k~Q_k \rho_f
Q_k^\dag
\end{equation}
and therefore $\rho\mapsto (\rho_f,\rho_s)$ is a bijective map. This
map is a sort of ``change of variables'' decoupling the slow part
from the fast part of the dynamics. Note that, in the slow part,
$\rho_s$, we have somehow removed the fast dynamics associated to
the optical state $\ket{e}$. Indeed, this change of variable leads to
a standard form:
\begin{alignat}{2}
&
\frac{d}{dt}\rho_f=-\frac{\Big(\sum_{k=1}^N\overline\Gamma_k\Big)}{2\epsilon}(\rho_f+P
\rho_f P)\notag\\
&\qquad\qquad\qquad-\frac{\imath}{\hbar}(P[H,\rho]+[H,\rho]P-P[H,\rho]P), \label{eq:tikhfast}\\
&\imath\hbar\frac{d}{dt}\rho_s=(1-P)[H,\rho](1-P)\notag\\
&\qquad\qquad\qquad+\frac{1}{\Big(\sum_{k=1}^N
\overline\Gamma_k\Big)}\sum_{k=1}^N \overline\Gamma_k
Q_k[H,\rho]Q_k^\dag.\label{eq:tikhslow}
\end{alignat}
where $\frac{1}{\epsilon}$ only appears  in  first equation defining $\dotex \rho_f$.
Therefore $\rho_f$ is associated to the fast part of the dynamics
and $\rho_s$ represents the slow part. The fast part is
asymptotically stable because $-\frac{\Big(\sum_{k=1}^N\overline
\Gamma_k\Big)}{2\epsilon}(\rho_f+P\rho_f P)$ defines a negative
definite super-operator on the space of Hermitian operators:
$$
\tr{-(\rho_f+P\rho_f P)\rho_f}=-(\|\rho_f\|^2+\|P\rho_f P\|^2).
$$
Moreover its inverse is given by :
\begin{equation}\label{eq:inv}
X\mapsto X-\frac{1}{2}P X P.
\end{equation}
Here we can apply the slow manifold
approximation~\eqref{eq:manifold} described in the
Appendix~\ref{sec:append}. Computing the first order terms, we find
the following approximation for $\rho_f$ with respect to $\rho_s$:
\begin{equation}\label{eq:order1}
\rho_f=\frac{-2\imath~\epsilon}{\hbar\Big(\sum_{k=1}^N
\overline\Gamma_k\Big)}~(P H \rho_s-\rho_s H P)+O(\epsilon^2).
\end{equation}
Inserting now the equations~\eqref{eq:inverse} into the
equation~\eqref{eq:tikhslow}, we have:\small
\begin{align*}
\frac{d}{dt}\rho_s &= -\frac{\imath}{\hbar}(1-P)[H,\rho_s](1-P)-\frac{\imath}{\hbar}(1-P)[H,\rho_f](1-P)\\
&+\frac{\imath}{\hbar\Big(\sum_{k=1}^N
\overline\Gamma_k\Big)}(1-P)\sum_{k=1}^N \overline \Gamma_k
[H,Q_k\rho_f Q_k^\dag] (1-P)\\
&-\frac{\imath}{\hbar\Big(\sum_{k=1}^N \overline\Gamma_k\Big)}
\sum_{k=1}^N \overline \Gamma_k~Q_k[H,\rho_f] Q_k^\dag,
\end{align*}\normalsize
where we have used~\eqref{eq:relation} and
\begin{equation}\label{eq:aux1}
Q_k\rho_s=\rho_s Q_k^\dag=0.
\end{equation}
Applying now the first order approximation~\eqref{eq:order1}, and
after some simple but tedious computations, we have
\begin{align}\label{eq:aux2}
\frac{d}{dt}\rho_s &=
-\frac{\imath}{\hbar}(1-P)[H,\rho_s](1-P)\notag\\
&-\frac{2\epsilon}{\hbar^2\Big(\sum_{k=1}^N
\overline\Gamma_k\Big)}\Big((1-P)HPH(1-P)\rho_s\notag\\
&\qquad\qquad\qquad\qquad\quad+\rho_s(1-P)HPH(1-P)\Big)\notag\\
&+\frac{4\epsilon}{\hbar^2\Big(\sum_{k=1}^N
\overline\Gamma_k\Big)^2}\sum_{k=1}^N \overline \Gamma_k Q_k H\rho_s
HQ_k^\dag+O(\epsilon^2).
\end{align}
Here, we have in particular applied~\eqref{eq:aux1} as well as the
fact that $Q_k P=Q_k$. Continuing  the computations, we have
\begin{multline}\label{eq:slow}
\frac{d}{dt}\rho_s=  -\frac{\imath}{\hbar}[\overline H,\rho_s]+
\\ 2\epsilon \sum_{k=1}^N \overline \Gamma_k\left(2\overline Q_k
\rho_s \overline Q_k^\dag -\overline Q_k^\dag \overline Q_k \rho_s
-\rho_s \overline Q_k^\dag \overline Q_k\right)
\end{multline}
where we have defined
\begin{equation}\label{eq:def1}
\overline H= (1-P) H (1-P)
\end{equation}
and
\begin{equation}\label{eq:def2}
\overline Q_k=\frac{1}{\hbar\Big(\sum_{k=1}^N
\overline\Gamma_k\Big)}(1-P)Q_k H (1-P).
\end{equation}
Note that
$$
\overline Q_k^\dag\overline Q_k=\frac{1}{\hbar^2\Big(\sum_{k=1}^N
\overline\Gamma_k\Big)^2}(1-P)HPH(1-P),
$$
which, in particular, allows us passing from~\eqref{eq:aux2}
to~\eqref{eq:slow}.

The situation is different for the measured output $y$. We have:
\begin{multline*}
y(t)=\frac{\Big(\sum_{k=1}^N \overline\Gamma_k\Big)}{\epsilon}
\tr{P\rho}=\frac{\Big(\sum_{k=1}^N \overline\Gamma_k\Big)}{\epsilon}
\tr{P\rho_f}\\ =\frac{-2\imath}{\hbar}~ \tr{P(P H \rho_s-\rho_s H
P)}+O(\epsilon).
\end{multline*}
But $\tr{P(PH\rho_s-\rho_s H P)}=0$. We should therefore consider
the second order terms otherwise the first order approximation yields  $y=0$. Using the Appendix~\ref{sec:append}, simple
but tedious computations end up  by the following natural
approximation:
\begin{equation}\label{eq:measure-2order}
y(t)=4\epsilon\Big(\sum_{k=1}^N \overline\Gamma_k\Big)\tr{\overline
P \rho_s} + O(\epsilon^2),
\end{equation}
where we have defined
$$
\overline P=\overline Q_k^\dag \overline
Q_k=\frac{1}{\hbar^2\Big(\sum_{k=1}^N
\overline\Gamma_k\Big)^2}(1-P)HPH(1-P).
$$
In order to derive~\eqref{eq:measure-2order}, we only need to
apply~\eqref{eq:singorder2} with the appropriate values of the
functions given in~\eqref{eq:tikhfast} and~\eqref{eq:tikhslow} and
the inverse map given in~\eqref{eq:inv}.

We can therefore prove the following theorem:
\begin{thm}\label{thm:main}
Consider $\rho$ the solution of the Markovian master
equation~\eqref{eq:fastdecay} and $\rho_s$ the solution of the slow
master equation~\eqref{eq:slow} with~\eqref{eq:def1}
and~\eqref{eq:def2}. Assume for the initial states
$|\rho(0)-\rho_s(0)|=\sqrt{\tr{(\rho(0)-\rho_s(0))(\rho(0)-\rho_s(0))}}=O(\epsilon)$.
Then
$$|\rho(t)-\rho_s(t)|=\sqrt{\tr{(\rho(t)-\rho_s(t))(\rho(t)-\rho_s(t))}}=O(\epsilon)$$
on a time scale $t\sim 1/\epsilon$.

Moreover the output $y(t)$  of the system (given
by~\eqref{eq:measure}) may be written as
in~\eqref{eq:measure-2order}.
\end{thm}

\begin{rem}
Note that, the approximation of this theorem is  stronger than the
usual one   only ensuring an error of order  $O(\epsilon)$ on a
finite time $T$ rather than on a time scale of $t\sim T/\epsilon$.
This stronger  result  is due to the Hamiltonian structure of the
dominant part of the dynamics.
\end{rem}

\begin{proof}
Applying~\eqref{eq:inverse} and the singular perturbation theory of
the appendix~\ref{sec:append}, we have
$\rho(t)=\widetilde\rho_s(t)+O(\epsilon)$ where
\begin{align}\label{eq:slowaux}
\frac{d}{dt}\widetilde\rho_s &=  -\frac{\imath}{\hbar}[\overline
H,\widetilde\rho_s]+ \notag\\ &2\epsilon \sum_{k=1}^N \overline
\Gamma_k\left(2\overline Q_k \widetilde\rho_s \overline Q_k^\dag
-\overline Q_k^\dag \overline Q_k \widetilde\rho_s
-\widetilde\rho_s \overline Q_k^\dag \overline Q_k\right)+O(\epsilon^2),\notag\\
\widetilde\rho_s(0)&=\rho(0).
\end{align}
Denoting by $\overset{\frown}{\delta\rho_s}= \widetilde\rho_s
-\rho_s$, we have
\begin{multline*}
\frac{d}{dt}\tr{\overset{\frown}{\delta\rho_s}^2}\leq\\
8\epsilon\sum_{k=1}^N \overline{\Gamma}_k
\left(\tr{Q_k\overset{\frown}{\delta\rho_s} Q_k^\dag
\overset{\frown}{\delta\rho_s}}-\tr{Q_k^\dag Q_k
\overset{\frown}{\delta\rho_s}^2}\right)\\
+\tr{O(\epsilon^2)\overset{\frown}{\delta\rho_s}}.
\end{multline*}
This, together with Cauchy-Schwartz inequality, implies
\begin{multline*}
\tr{\overset{\frown}{\delta\rho_s}^2(t)}\leq
\tr{\overset{\frown}{\delta\rho_s}^2(0)}+\epsilon L \int_0^t
\text{Tr}^{\frac{1}{2}}\left[\overset{\frown}{\delta\rho_s}^4(\tau)\right]d\tau\\+\epsilon^2
C \int_0^t
\text{Tr}^{\frac{1}{2}}\left[\overset{\frown}{\delta\rho_s}^2(\tau)\right]d\tau,
\end{multline*}
where $L$ and $C$ are positive constants. Note that,
$\overset{\frown}{\delta\rho_s}^2$ being definite positive, we have
$$
\text{Tr}^{\frac{1}{2}}\left[\overset{\frown}{\delta\rho_s}^4(\tau)\right]\leq
\text{Tr}\left[\overset{\frown}{\delta\rho_s}^2(\tau)\right].
$$
Therefore, noting
$\xi=\sqrt{\text{Tr}\left[\overset{\frown}{\delta\rho_s}^2(\tau)\right]}$,
we have
$$
\xi^2(t)\leq \xi^2(0)+\epsilon L \int_0^t
\xi^2(\tau)d\tau+\epsilon^2 C \int_0^t\xi(\tau)d\tau.
$$
Denoting $\zeta=\xi(t)+\frac{C}{2L}\epsilon$, some simple
computations lead to
\begin{multline*}
\zeta^2(t)\leq 2\xi^2(0)+2\epsilon L\int_0^t
\zeta^2(\tau)d\tau-\frac{C^2}{2L}\epsilon^3~t+\frac{C^2}{2L^2}\epsilon^2\\
\leq \xi^2(0)+\frac{C^2}{2L^2}\epsilon^2+2\epsilon L\int_0^t
\zeta^2(\tau)d\tau.
\end{multline*}
Applying the Gronwall lemma, we have
$$
\zeta^2(t)\leq
\left[\xi^2(0)+\frac{C^2}{2L^2}\epsilon^2\right]e^{2\epsilon L t}.
$$
Noting that, by the Theorem's assumption, $ \xi(0)= O(\epsilon)$, we
have $\zeta(t)=O(\epsilon)$ on a time scale of $t\sim 1/\epsilon$.
As $\xi(t)=\zeta(t)+O(\epsilon)$, this trivially finishes the proof.
\end{proof}

From a practical point of view, the main result of this section is as follows.  The correct  slow approximation (also called by physicists adiabatic approximation) of the system  described by
\begin{multline*}
     \frac{d}{dt}\rho  =  -\frac{\imath}{\hbar}[ H,\rho]
 \\  + \sum_{k=1}^N \frac{\Gamma_k}{2} \left(2Q_{k}
\rho  Q_{k}^\dag - Q_{k}^\dag  Q_{k} \rho
-\rho  Q_{k}^\dag  Q_{k}\right)
\end{multline*}
with $Q_k=\ket{g_k}\bra{e}$ and  where the $\Gamma_k$'s are  much
larger than $\frac{H}{\hbar}$  and  where the output reads $y
=\sum_k \Gamma_k \tr{Q_k^\dag Q_k \rho} $,  is given by
\begin{align}
\frac{d}{dt}\rho_s& =  -\frac{\imath}{\hbar}[ H_s,\rho_s]
 \notag \\  &+\quad \sum_{k=1}^N \frac{4\Gamma_k}{2} \left(2Q_{s,k}
\rho_s  Q_{s,k}^\dag - Q_{s,k}^\dag  Q_{s,k} \rho_s
-\rho_s  Q_{s,k}^\dag  Q_{s,k}\right)
    \label{eq:slowbis}
\end{align}
where $\rho_s$ is the density operator associated to the space spanned by the $\ket{g_k}$'s, where   the slow Hamiltonian is $$
 H_s= (1-P) H (1-P) $$
and the slow jump operators are
$$
Q_{s,k}=Q_k \frac{H}{\hbar \Gamma}(1-P).
$$ Here, we have set $\Gamma=\sum_k \Gamma_k$ and $P=\ket
e\bra e$. The slow approximation of the measured output is still
given by the standard formula
$$
y_s= \sum_{k=1}^{n} 4 \Gamma_k \tr{Q_{s,k}^\dag Q_{s,k} \rho_s}
.
$$

\section{Physical interpretation}\label{sec:phys}
In this section, we provide a physical interpretation of the last
section's result for  the particular Hamiltonian of the
system~\eqref{eq:twoscale}. We get
$$
H_s = \sum_k \delta_k \ket{g_k}\bra{g_k}
$$
and
$$
Q_{s,k} = \frac{\sqrt{\sum_l |\Omega_l|^2}}{\sum_l \Gamma_l}  \ket{g_k} \bra{b_\Omega}
\quad \text{with}\quad \ket{b_\Omega} = \frac{\sum_l \Omega_l \ket{g_l}}{\sqrt{\sum_l |\Omega_l|^2}}
.
$$
Let us set
$$
\gamma_k =  \frac{{\sum_l |\Omega_l|^2}}{\left(\sum_l\Gamma_l\right)^2} ~ 4\Gamma_k
$$
the slow master equation reads
\begin{multline*}
    \dotex \rho_s =
    -\imath \left[\sum_k \delta_k \ket{g_k}\bra{g_k}~,~ \rho_s \right]
    \\
    + \sum_k \frac{\gamma_k}{2}
    \text{\Large (}2\bket{{b_\Omega}|\rho|{b_\Omega}} \ket{g_k}\bra{g_k} -\ket{b_\Omega}\bra{b_\Omega}\rho-\rho \ket{b_\Omega}\bra{b_\Omega}
  \text{\Large )}
\end{multline*}
with $y_s = \left(\sum_k\gamma_k\right)
\bket{b_\Omega|\rho_s|b_\Omega}$. Thus, whenever all the de-tunings
$\delta_k$ vanish, the unitary state $\ket{b_\Omega}$ is the bright
state and the vector-space orthogonal to $\ket{b_\Omega}$ is a
decoherence free space since on this sub-space, the
Lindbald-Kossakowski terms identically vanish and the output $y$ is
null. Notice that  the controls $\Omega_k$  appear  only in the
decoherence terms and have disappeared form the slow Hamiltonian.

If we  restrict ourselves to the case of a 3-level $\Lambda$-system ($N=2$),
 we have  $H_s=\frac{\delta}{2}(\ket{g_1}\bra{g_1}-\ket{g_2}\bra{g_2})$,
$$
\ket{b_\Omega}=\frac{\Omega_1}{\sqrt{|\Omega_1|^2+|\Omega_2|^2}}
\ket{g_1}+\frac{\Omega_2}{\sqrt{|\Omega_1|^2+|\Omega_2|^2}}
\ket{g_2}
$$
is the bright state of the $\Lambda$-system, as in the context of
the coherent population trapping. As it can be seen easily, whenever
no de-tuning is admitted ($\delta=0$), the dark state
$$
\ket{d}=\ket{b}^\perp=\frac{\Omega_2^*}{\sqrt{|\Omega_1|^2+|\Omega_2|^2}}
\ket{g_1}-\frac{\Omega_1^*}{\sqrt{|\Omega_1|^2+|\Omega_2|^2}}
\ket{g_2}
$$
is the only equilibrium state of the slow dynamics.

\section{Numerical simulations}\label{sec:num}
\begin{figure}[h]
\begin{center}
\flushleft\includegraphics[width=0.52\textwidth, height=.25
\textwidth]{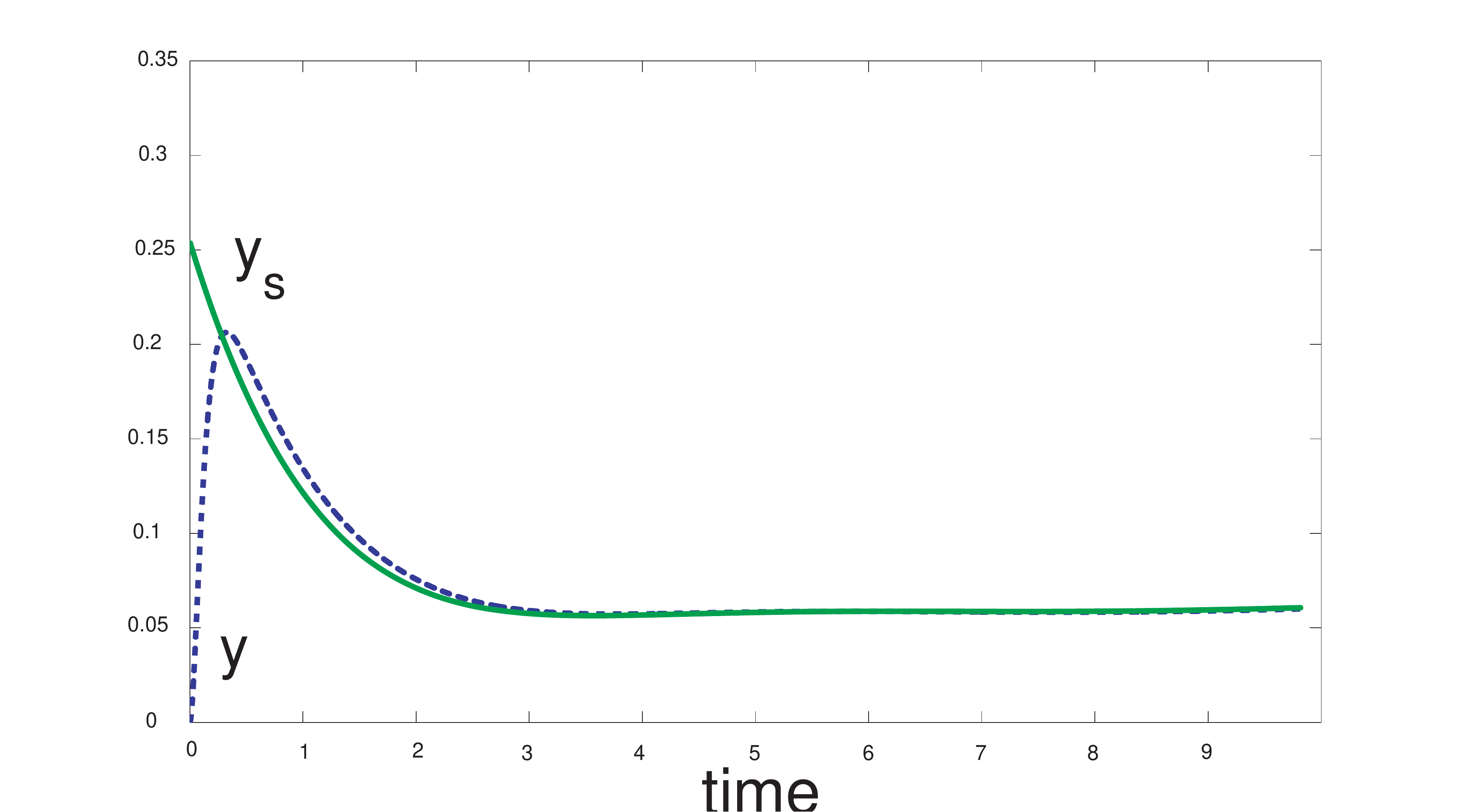} \caption{The output of the reduced slow
dynamics~\eqref{eq:slowbis} (solid line) versus the  output for the
(4+1)-level system~\eqref{eq:twoscale} (dashed line). The parameters
are listed in~\eqref{eq:4level}.  Even if  the time-scale separation
is not so large with an $\epsilon$ around $\frac{1}{4}$, such
adiabatic approximation captures quite precisely the slow part of
the dynamics, as stated by theorem~\ref{thm:main}.
}\label{fig:slowfast}
\end{center}
\end{figure}

Finally let us check the relevance of the adiabatic reduction result
of the Section~\ref{sec:phys} in a simulation. Here, we consider a
(4+1)-level system given by the following parameters
\begin{align}\label{eq:4level}
\delta_1 &=0.5\quad \delta_2=1.2\quad \delta_3=0.7\quad
\delta_4=1.0\notag\\
\Omega_1 &=1.0\quad \Omega_2=1.2\quad \Omega_3=1.1\quad
\Omega_4=1.3\notag\\
\Gamma_1 &=5.0\quad \Gamma_2=4.0\quad
\Gamma_3=7.0\quad \Gamma_4=5.0
\end{align}
The simulations of Figure~\ref{fig:slowfast} illustrate the output
signal derived from the reduced slow dynamics, $y_s$, versus the slow/fast dynamics,
 $y$. The simulation time is taken to be $2.5T_s$, where
$T_s$, the time scale of the slow system is given by
$$
T_s=\frac{ \Gamma_1+ \Gamma_2+
\Gamma_3+
\Gamma_4}{\Omega_1^2+\Omega_2^2+\Omega_3^2+\Omega_4^2}.
$$
The initial conditions  are identical   $\rho(0)=\rho_s(0)= \frac{\sum_{k=1}^{4} \ket{g_k}\bra{g_k}}{4}$. We observe, for the slow/fast dynamics, a first fast transient corresponding to the relaxation time   of the fast dynamics  for $t$ between $0$ and $1/\Gamma_k$, i. e. $t\in[0,1/4]$, and then two slow transients very similar for  both master equations: the slow approximation is clearly valid only for time-scale much larger than $1/\Gamma_k$.

\section*{Conclusion}
We observed that for an ensemble of independent and identical
quantum systems, and whenever the decoherence dynamics due to the
measurement is much faster than the other dynamics, the adiabatic
approximation helps us to find the slow dynamics as well as the
measurement result with respect to the slow dynamics. Note that in
this new system, the decoherence term can be removed in a first
order approximation. We obtain therefore a system of the form
$$
\frac{d}{dt}\rho_s=-\frac{\imath}{\hbar}[H_s,\rho_s],
$$
where the control appears linearly in the reduced Hamiltonian $H_s$.
This system corresponds to a bilinear system with the wavefunctions
as state variables. Furthermore, we have access to a measurement $y$
given by the reduced slow evolution.  We can henceforth consider a
control problem with continuous measurement associated to this
system.

Notice that,
by some simple but tedious computations, one can
extend the result of this paper to the more general case of an
$N+M$-level system with $N$ metastable ground states  and $M$ highly
unstable excited states.

\appendix
\section{Singular perturbation theory}\label{sec:append}
This appendix has for goal to remind an approximation technique that
can be perfectly justified using the geometrical tools of singular
perturbation and the center
manifold~\cite{khalil-book,carr-book,duchene-rouchon-ces96}.

Consider the slow/fast system ($x$ and $y$ are of arbitrary
dimensions, $f$ and $g$ are regular functions)
$$
\dotex x=f(x,y),\qquad \dotex y=-\frac{1}{\epsilon}Ay+g(x,y)
$$
where $x$ and $y$ are respectively the slow and fast states
(Tikhonov coordinates), all the eigenvalues of the matrix $A$ have
strictly positive real parts, and $\epsilon$ is small strictly
positive parameter. Therefore the invariant attractive manifold
admits for  equation
\begin{equation}\label{eq:manifold}
y=\epsilon A^{-1}g(x,0)+O(\epsilon^2)
\end{equation}
and the restriction of the dynamics on this slow invariant manifold
reads
\begin{multline*}
\dotex x=f(x,\epsilon
A^{-1}g(x,0))+O(\epsilon^2)\\
=f(x,0)+\epsilon\frac{\partial f}{\partial
y}|_{(x,0)}A^{-1}g(x,0)+O(\epsilon^2).
\end{multline*}
The Taylor expansion of $g$ can be used to find the higher order
terms. For example, the second order term in the expansion of $y$ is
given by:
\begin{multline}\label{eq:singorder2}
y=\epsilon A^{-1}g(x,0) +\\
\epsilon^2 A^{-1}\left(\frac{\partial g}{\partial
y}|_{(x,0)}A^{-1}g(x,0)-A^{-1}\frac{\partial g}{\partial
x}|_{(x,0)}f(x,0)\right)+O(\epsilon^3),
\end{multline}
and so on.


\end{document}